\documentclass[conference]{IEEEtran}
\usepackage{xcolor}

\usepackage{cite}
\usepackage{amsmath,amssymb,amsfonts}
\usepackage{graphicx}
\usepackage{textcomp}
\usepackage{booktabs}
\usepackage{multirow}
\usepackage{multicol}
\usepackage{bm}
\usepackage{hyperref}
\usepackage{url}
\usepackage{subfig}
\usepackage[normalem]{ulem}
\usepackage{threeparttable}
\usepackage{verbatim}
\usepackage{xcolor}
\definecolor{darkgreen}{rgb}{0,0.8,0}

\IEEEoverridecommandlockouts                              

\title{\LARGE \bf
Short-Term Load Forecasting for AI-Data Center}

\author{\IEEEauthorblockN{Mariam Mughees, Yuzhuo Li, Yize Chen, and Yunwei Ryan Li\IEEEauthorrefmark{1}
}

    \IEEEauthorblockA{\IEEEauthorrefmark{1} Department of Electrical and Computer Engineering\\ University of Alberta, Edmonton, Alberta, Canada 
    \\\{mughees, yuzhuo, yize.chen, yunwei.li\}@ualberta.ca}}

\begin{document}

\maketitle

\begin{abstract}
Recent research shows large-scale AI-centric data centers could experience rapid fluctuations in power demand due to varying computation loads, such as sudden spikes from inference or interruption of training large language models (LLMs). As a consequence, such huge and fluctuating power demand  pose significant challenges to both data center and power utility operation. Accurate short-term power forecasting allows data centers and utilities to dynamically allocate resources and power large computing clusters as required. However, due to the complex data center power usage patterns and the black-box nature of the underlying AI algorithms running in data centers, explicit modeling of AI-data center is quite challenging. Alternatively, to deal with this emerging load forecasting problem, we propose a data-driven workflow to model and predict the short-term electricity load in an AI-data center, and such workflow is compatible with learning-based algorithms such as LSTM, GRU, 1D-CNN. We validate our framework, which achieves decent accuracy on data center GPU short-term power consumption. This provides opportunity for improved power management and sustainable data center operations. 

\end{abstract}

\section{INTRODUCTION}
The data center industry is expanding rapidly, driven by increasing cloud services, Artificial Intelligence (AI)/Machine Learning (ML) advancements, and data storage needs. Global AI-related electricity demand is projected to grow significantly due to technological, economic, and social factors~\cite{lin2023adapting}. This decade has seen major tech companies, like Google and Oracle, invest heavily in new and expanded data center facilities globally \cite{crn2024, dcd2024}. However, these growing demands pose challenges for power grids, leading to concerns about whether they can handle these novel, high-power-density loads, as highlighted in PJM’s report on congested lines and rising electricity costs\cite{PJM}. Additionally, environmental considerations push data centers toward modular designs and renewable energy. 
AI-focused data centers, with unprecedented per-rack power densities, introduce significant power grid transients, akin to those from Electric Vehicles (EVs) and renewable energy sources \cite{lin2024exploding}.

\begin{figure}[h]
    \centering
    \includegraphics[width=0.9\linewidth]{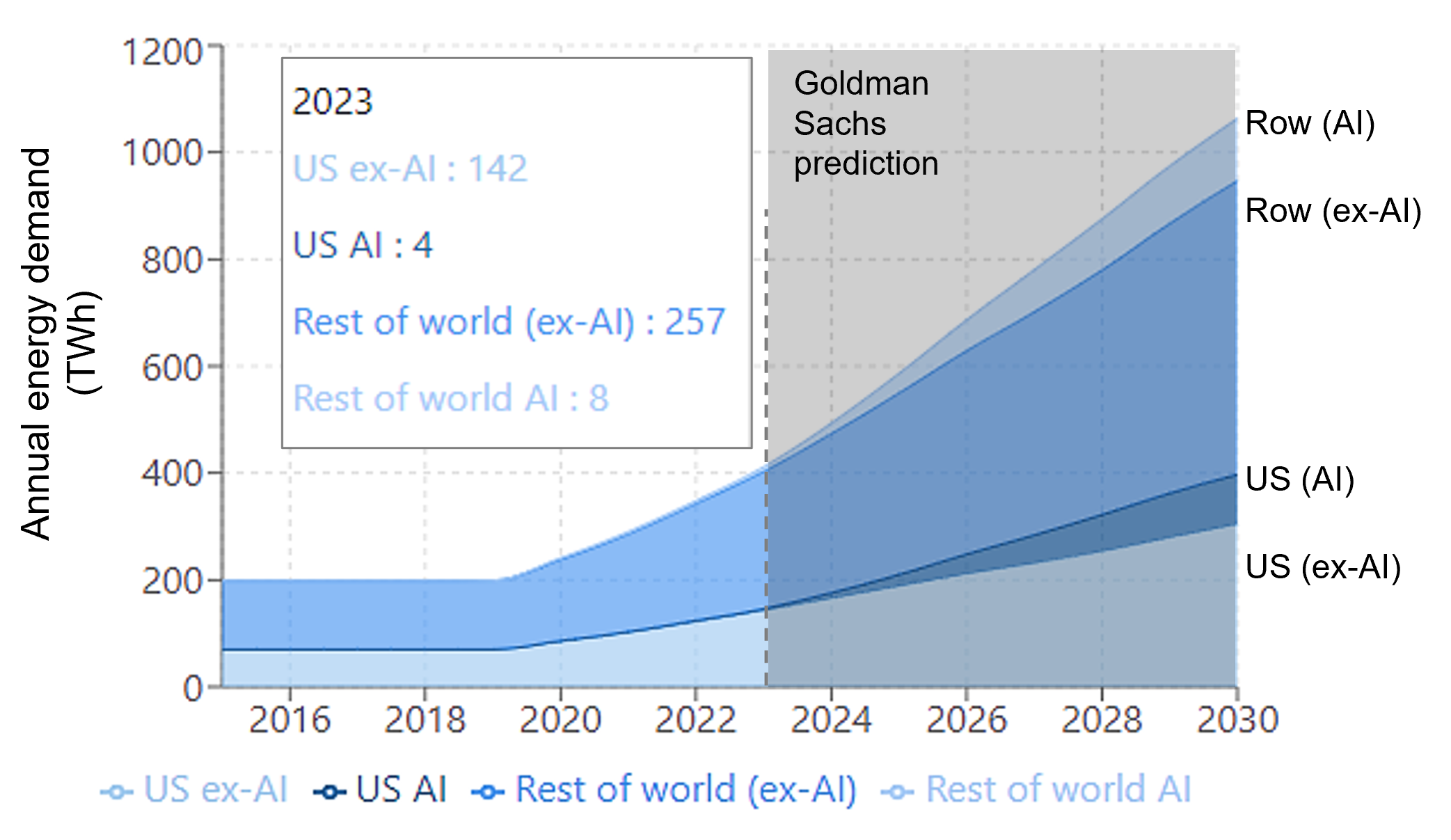}
    \caption{The stack-area diagram of data center annual energy demand  (adapted from \cite{GoldmanSachs2024}). (ex-AI: excluding AI's demand.)}
    \label{fig:AI-power}
\end{figure}

Data center load can introduce huge and instant power variations during data center cold start, shutdown, load shifting or sudden interruption of load. As these large data centers are consuming tens to hundreds of MW power, and may add up to several MW of power changes in few seconds, this can affect grid’s frequency control and may demand more responsive frequency regulations in the system. 
Recently, data centers also operate many power electronics equipment such UPS, control switches and rectifiers. During a transient event, these non-linear components introduce harmonics into the grid and reduce power quality.  Thus to improve grid reliability, forecasting the energy consumption is necessary for control strategies such as Automatic Generation Control (AGC), particularly when handling very large power transients caused by AI loads, harmonics and frequency variations~\cite{li2024unseen}. Predicting sags and swells upcoming in power consumption can also help AGC to prepare for these fast ramp-up events.

Existing literature \cite{amvrosiadis2018diversity, blocher2021switches, Wilkins2024HybridClusters} consider different aspects such as workloads and failure rates,  and these features are energy-related. However even diverse datasets with unidentified jobs make predictions difficult to implement and \cite{Wang2024UtilizationPrediction} also highlights opportunities like reinforcement learning to improve scheduling.
\cite{rossi2015forecasting},\cite{meisner2009powernap} discuss Graphic Processing Unit (GPU) energy management by considering active and idle status according to usage and also based on prediction. 
\cite{shoukourian2020forecasting} considers regression techniques such as Auto Regressive Integrated Moving Average (ARIMA) and fault tree to predict power and failure events in data center facilities. \cite{Wilkins2024HybridClusters} highlights research related to user behavior and how user interaction can affect power.
A convolution neural network (CNN) based technique is presented in \cite{Bai2022DNNAbacus} which considers GPU workloads power consumption, especially by large language models (LLMs) and provides better results than ARIMA.
A Deep Neural Network (DNN) is proposed to predict the computational cost of LLM model training in cloud \cite{Patel2024PowerLLMs}. 
Short-term data center power forecasting is important specifically for the dynamic and resource-intensive nature of AI and for high-performance computing workloads~\cite{Hu2021GPUDatacenters}. Traditional models like ARIMA struggle with complex patterns in high-dimensional data. In contrast, adopting DNNs such as LSTM, GRU, and 1D-CNN can excel in forecasting power consumption for multivariate time series.

This is evident from the literature that data center power consumption is a complex and critical task and depends on many factors such as equipment installed, workload types and operating conditions. As collected data is very versatile in nature and has many sudden dips and peaks which makes forecasting an important and critical task. 
In this work, we address the challenge of the lack of quality datasets for analyzing energy-intensive GPU-based AI workloads, particularly for training large language models (LLM).
We design a general workflow using LSTM, GRU, and 1D-CNN architectures trained on the MIT Supercloud dataset, capturing detailed GPU power metrics over 8 months with a 1-second granularity and a 300-
look back window predicts power consumption 90 seconds ahead.
Results in the manuscript will show these models achieve decent prediction accuracy, validating their effectiveness for short-term power forecasting in AI-data centers.


\section{Data Center Load Forecasting}
\subsection{Problem Formulation}
\begin{figure}
    \centering
    \includegraphics[width=0.95\linewidth]{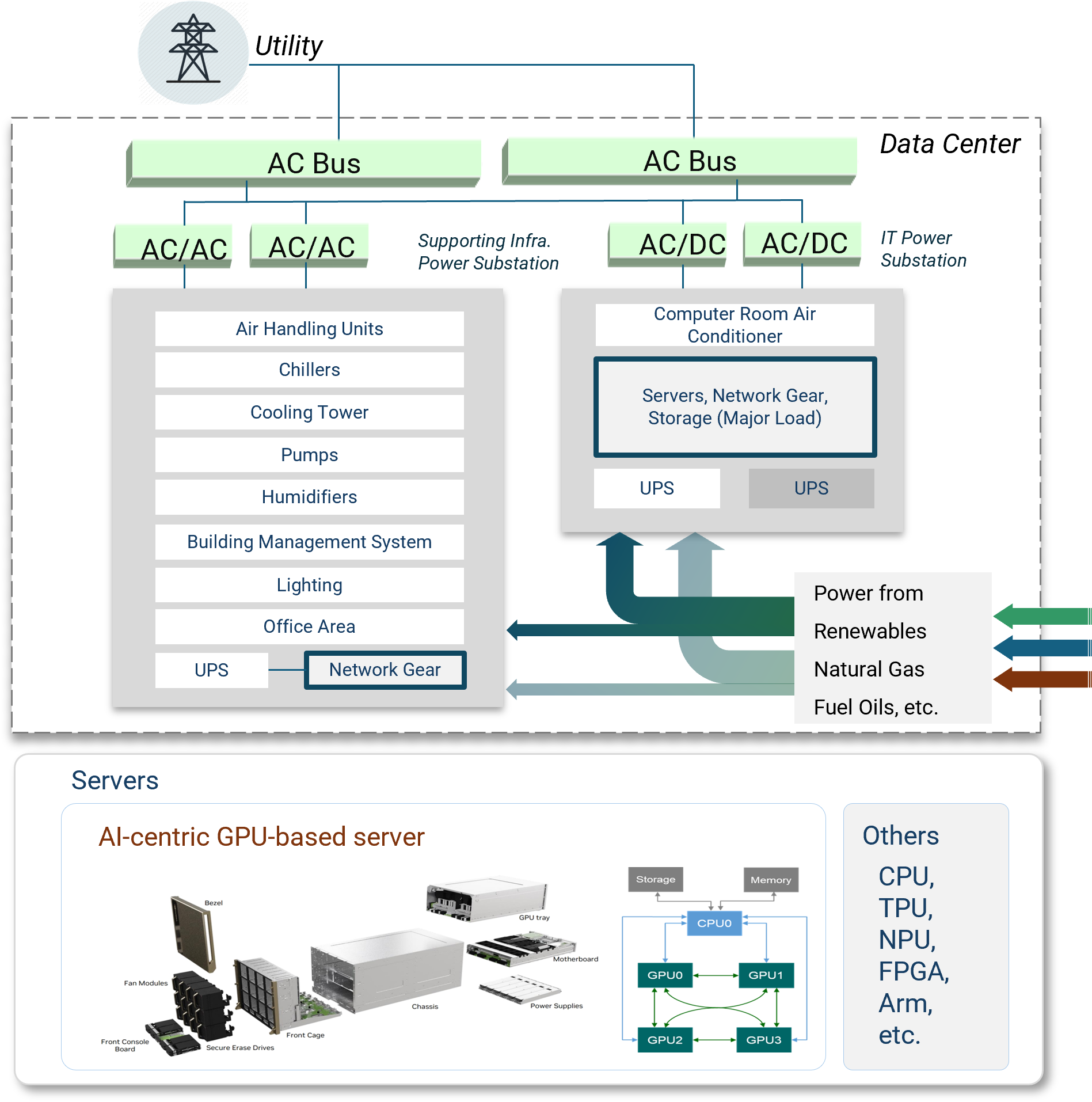}
    \caption{The electricity demand of AI-centric data center\cite{li2024unseen}.}
    \label{fig:AI-datacenter}   
\end{figure}

The rapid growth of AI computing has transformed data center power requirements. Modern AI workloads feature higher power densities (300W–1,200W per GPU), rapid power fluctuations (e.g., $>$132 kW/s at the rack level with NVIDIA GB200 NVL72 \cite{schneider2024ai}), and complex, non-linear scaling behaviors. Accurately forecasting these dynamics is crucial for infrastructure design, operational stability, cost optimization, and capacity planning to meet the rising demands of AI workloads. A typical data center is depicted in Fig.\ref{fig:AI-datacenter}, where computing infrastructure and supporting facilities are main loads, highlighting the complex power infrastructure necessary for an AI-centric data center.

\textbf{Defining the Time Series Data: }
Let \( X(t) \) represent the load at time step \( t \). The input data for each forecasting point will include the load values from the previous $H$ steps, since the future load is related to the short-term history load recorded. This means that for a given time step \( t \), the input sequence \( \mathbf{X}_{h}\) is defined as:

\begin{equation}
\mathbf{X}_{h} = \{ X(t-H), X(t-H+1), \dots, X(t-1) \}.
\end{equation}
Here, $\mathbf{X}_{h}$ is a vector of load values of $H$ elements, representing the load history up to time \( H-1 \).
The goal is to predict the load values for the next 90 steps based on the input sequence \( X(t) \). At time step $t$, define the forecasting sequence \( Y_f \) as:
\begin{equation}
\mathbf{Y}_p = \{ Y(t), Y(t+1), \dots, Y(t+P)\}.      
\end{equation}
In the above formulation, \( \mathbf{Y}_p\) denotes a vector of predicted load values over the next $P$ predicted time steps.
Let \( f(\cdot) \) represent the forecasting function (e.g., a neural network) that maps the input sequence \( \mathbf{X}_{h} \) to the output sequence \( \mathbf{Y}_p\):
\begin{equation}
    \mathbf{Y}_p=f(\mathbf{X}_{h} ).
\end{equation}

In real-world systems, load patterns are influenced by many different factors such as environmental factors, user behaviors, and demographics of infrastructure. This makes load forecasting a challenging task. As in this study data-driven forecasting techniques employed which help model to learn dynamics and nonlinearity from previous data. To fulfill this purpose, the number of look-back and forecasting horizon are defined and utilized.

To train the data center load forecasting model \( f \), we can minimize the mean squared error (MSE) over the forecasting horizon, given by:
\begin{equation}
\text{MSE} = \frac{1}{P} \sum_{i=0}^{P} \left( Y(t+i) - \hat{Y}(t+i) \right)^2;
\end{equation}
where \( \hat{Y}(t+i) \) represents the predicted load at time \( t+i \), and \( Y(t+i) \) represents the actual load. \(P\) is the length of forecasting horizon.

\begin{figure*}
    \centering
    \includegraphics[width=0.95\linewidth]{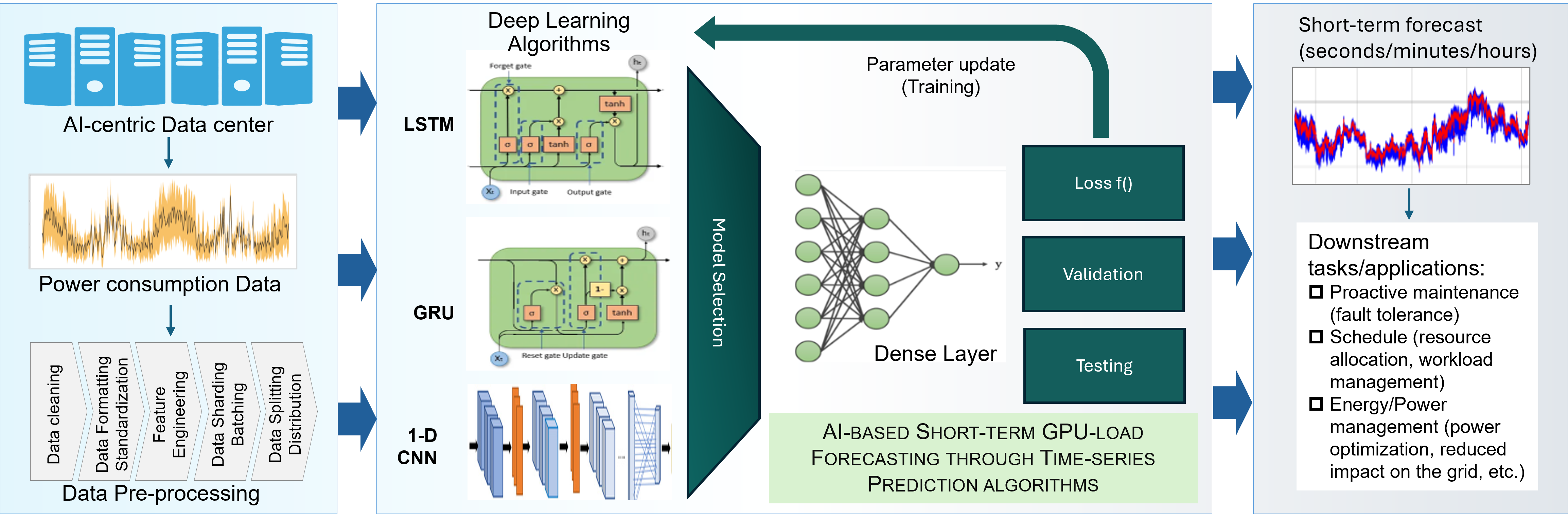}
    \caption{The workflow of the AI-based short-term forecasting of AI-data center through time-series prediction algorithms}
    \label{fig:workflow}    
\end{figure*}

\subsection{Short-Term Data-Center  GPU-Power Forecasting}
Near to real-time forecasting can help achieve improvements in many folds such as grid, cost reduction and load management. 
To achieve this forecasting goal, we design and follow a three-stage workflow. Fig.\ref{fig:workflow} outlines our proposed AI-based workflow for short-term GPU load forecasting through time-series prediction algorithms, divided into three main stages:

\subsubsection{Data Collection and Pre-Processing}

The data can be captured through either the hardware or software-based way. 
Hardware capture can be done through power monitor devices by sensing the power cords of the GPU and CPU motherboard or Power Supply Unit inside the rack.
For software capture, the computing unit is based on GPU, CPU or others as mentioned in Fig.\ref{fig:AI-datacenter}, then, vendor-specific commands can be used to facilitate the collection process. For instance, \texttt{nvidia-smi}
command can be considered to measure Nvidia's GPU power consumption. Collected data will include GPU power consumption (constitutes the majority of the total power consumption in AI-centric workloads), memory utilization, GPU temperature, and storage.
After the data collection, raw data undergoes several pre-processing steps, including pre-processing, Min-Max normalization, data slicing, etc., to prepare it for feeding into deep learning models.

\subsubsection{ Model Training}

Different deep learning architectures, e.g., LSTM, GRU, and 1-D CNN, can be deployed for time-series forecasting. These three models are selected due to their superior capabilities in dealing with sequential data. The model undergoes training, parameter updates, and performance validation to optimize accuracy.

\subsubsection{Forecasting and Application}

The model outputs short-term forecasts of different lengths of periods in advance (e.g., from seconds to minutes) based on the target various downstream applications, such as proactive maintenance, resource scheduling, and energy/power management. 

\subsection{Time-Series Models}

There are quite a few AI/ML methods that have been proposed for time-series prediction tasks, while DNNs are getting significant importance in fields of time series forecasting as these networks are able to grab complex details from temporal patterns and have multiple layers hidden between input and output. 
Algorithms like Recurrent neural networks (RNNs)  use back propagation through time (BPTT) which helps  memorize and analyze information from past time series. RNNs are used for continuous data and are very powerful for capturing dynamics of sequence data. However, these methods suffer from problem of vanishing or exploding gradients when trained on very long data sequences.
To handle such issues, idea of an explicit memory augmentation is being implemented in practice in LSTM network. The specifically designed memory cell functions as gated leaky neuron, which has a self- connection to itself at next step and has unity weight, so it duplicates its own value and adds the external signal. And this self-connection is multiplicatively gated by another unit which decides when to clear memory \cite{ye2024deep, bengio2013advances}.

\section{Numerical Simulations}

\subsection{Data Pre-processing on Real-world Data Center GPU-Load Dataset}
\begin{figure*}
    \centering
    \includegraphics[width=0.85\linewidth]{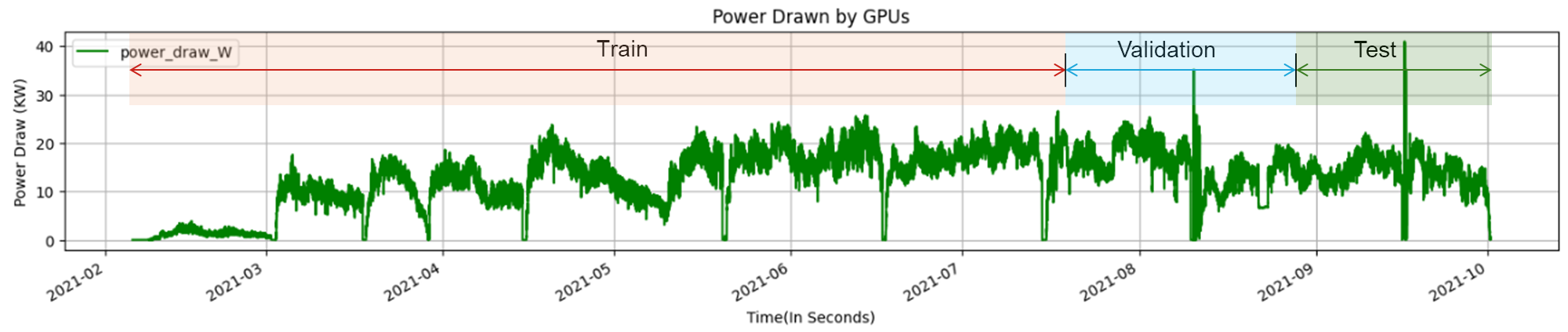}
    \caption{MIT Supercloud Dataset used in the simulation and the data split.}
    \label{fig:split}    
\end{figure*}

In this study, we address the data center power consumption forecasting problem using a real-world dataset from the MIT Supercloud \cite{samsi2021supercloud}, a high-performance computing (HPC) system (GPU: Nvidia Volta V100, CPU: Intel Xeon Gold 6248). The dataset spans February to October 2021 and includes 100-millisecond interval logs of GPU/CPU utilization, scheduling details, and physical parameters like temperature. Key GPU metrics include power, memory, utilization, and temperature, with anonymized user data organized by job ID and node. Aggregated GPU power consumption peaks at 45 kW across 448 GPUs.
The dataset details workload composition, dominated by vision networks (e.g., U-Net: 1,431 jobs; VGG, ResNet, and Inception follow), language models (e.g., BERT: 189 jobs; DistillBERT: 172 jobs), and graph neural networks (e.g., SchNet, DimeNet: lower counts). Pre-processing maintains a 1-second granularity, with power consumption aggregated by job ID and node to reflect total power drawn from the local distribution system. After normalization via a min-max scaler, the data uses a 300-lookback window to predict 90 seconds ahead. Fig.\ref{fig:split} illustrates the GPU power consumption trends and train-validation-test splits (ratios: 0.7, 0.15, 0.15).

\subsection{Simulation Results}

\begin{figure*}
    \centering
    \includegraphics[width=0.95\linewidth]{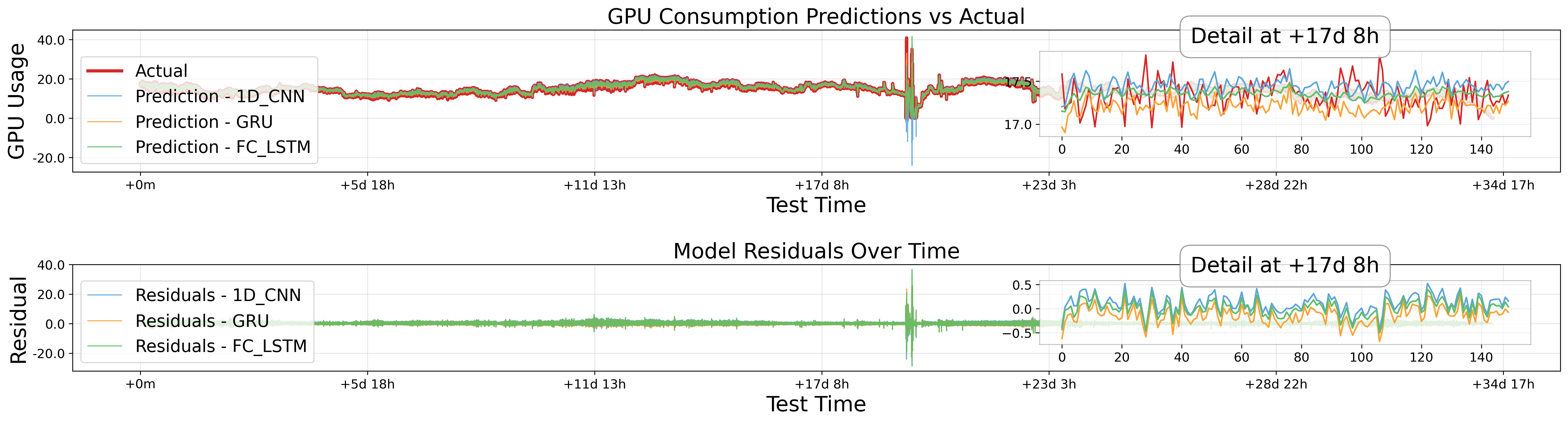}
    \caption{Prediction results (upper) and prediction error in terms of residuals (lower) for 1D\_CNN, GRU, and LSTM. Zoom-in view is visualized to the right.}
    \label{fig:full}    
\end{figure*}

\begin{figure}
    \centering
    \includegraphics[width=0.9\columnwidth]{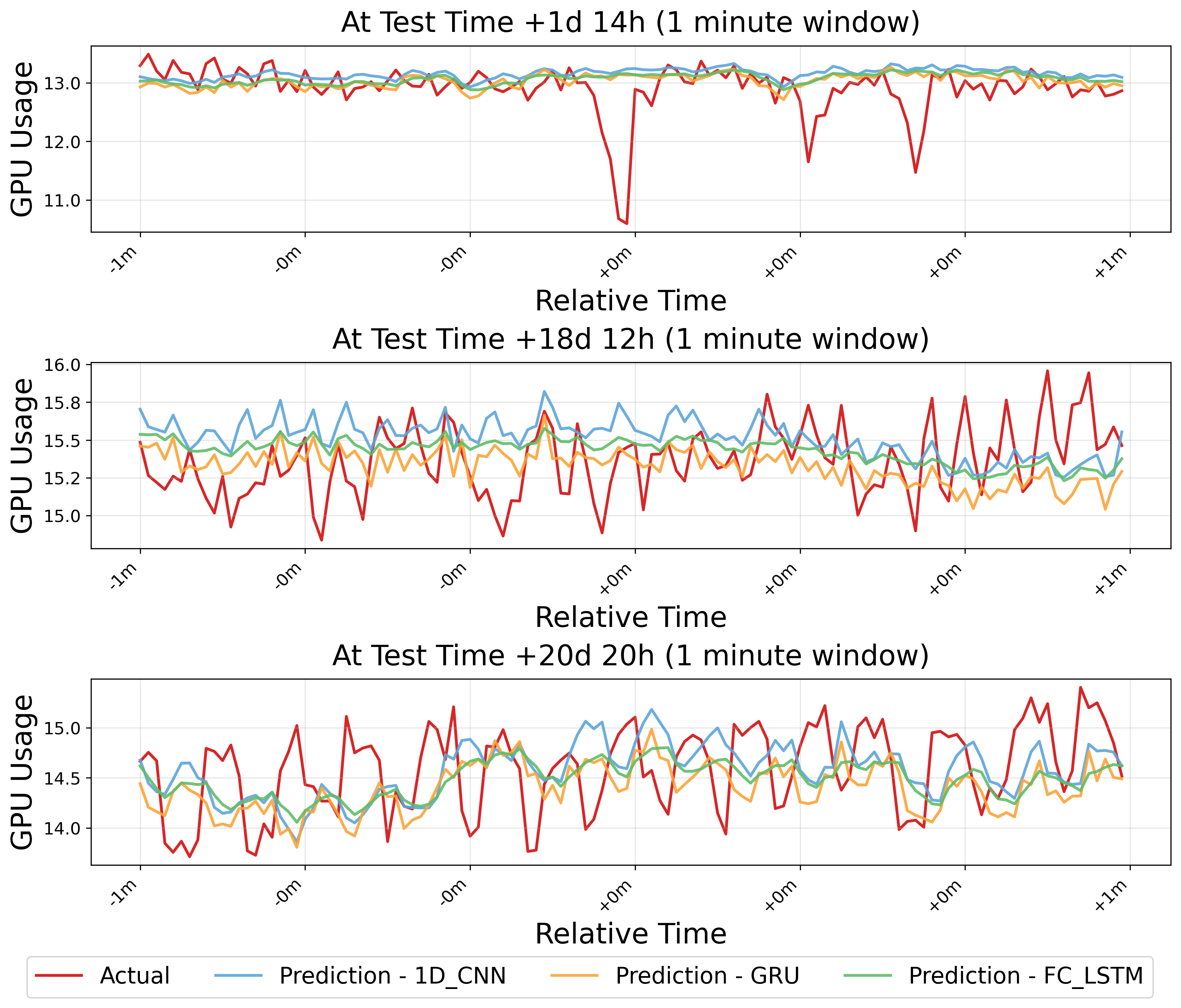}
    \caption{Prediction results (1 minute).}
    \label{fig:1_minute}    
\end{figure}

\begin{figure}
    \centering
    \includegraphics[width=0.9\columnwidth]{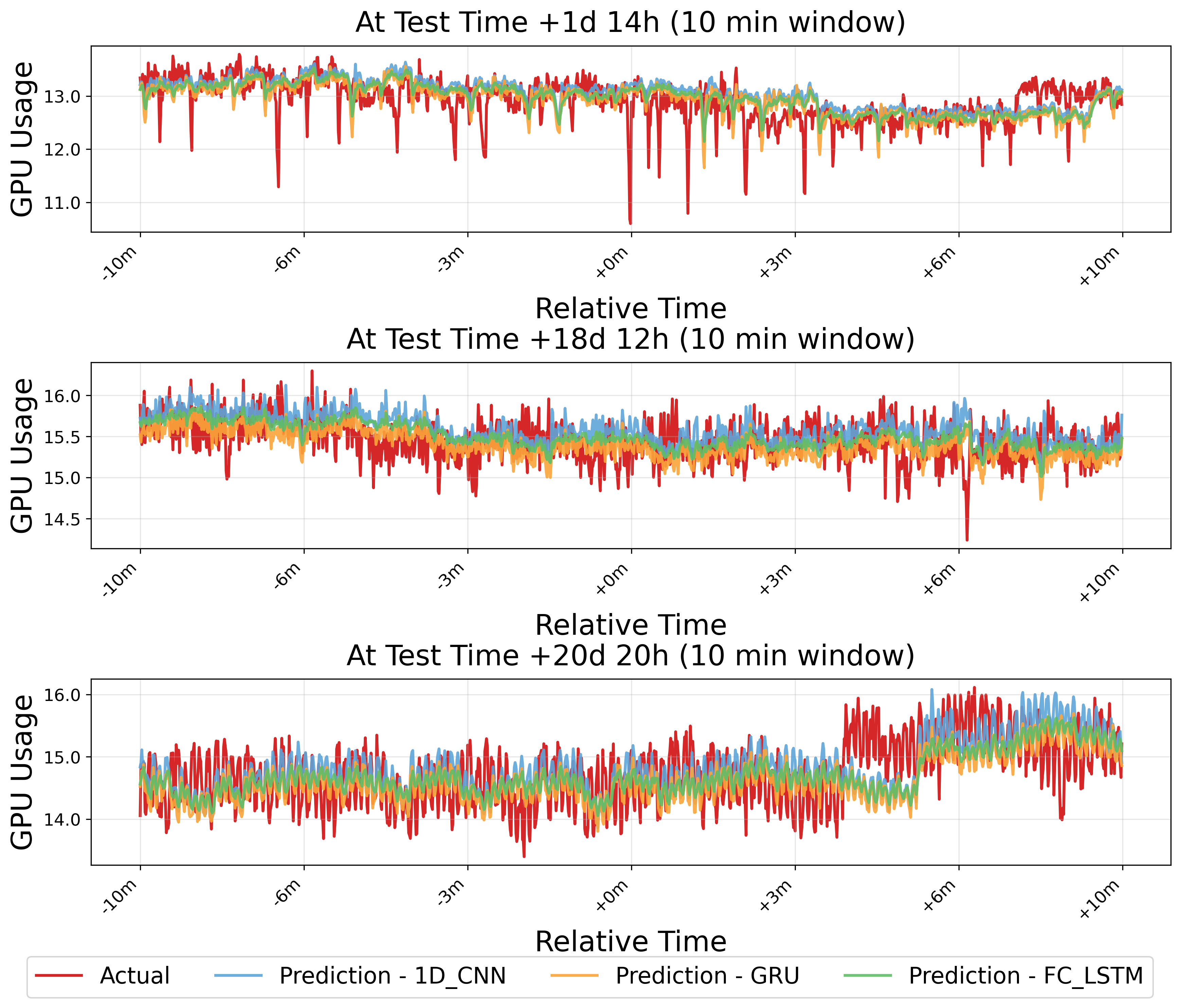}
    \caption{Prediction results (10 minutes).}
    \label{fig:10_min}    
\end{figure}

\begin{figure}
    \centering
    \includegraphics[width=0.9\columnwidth]{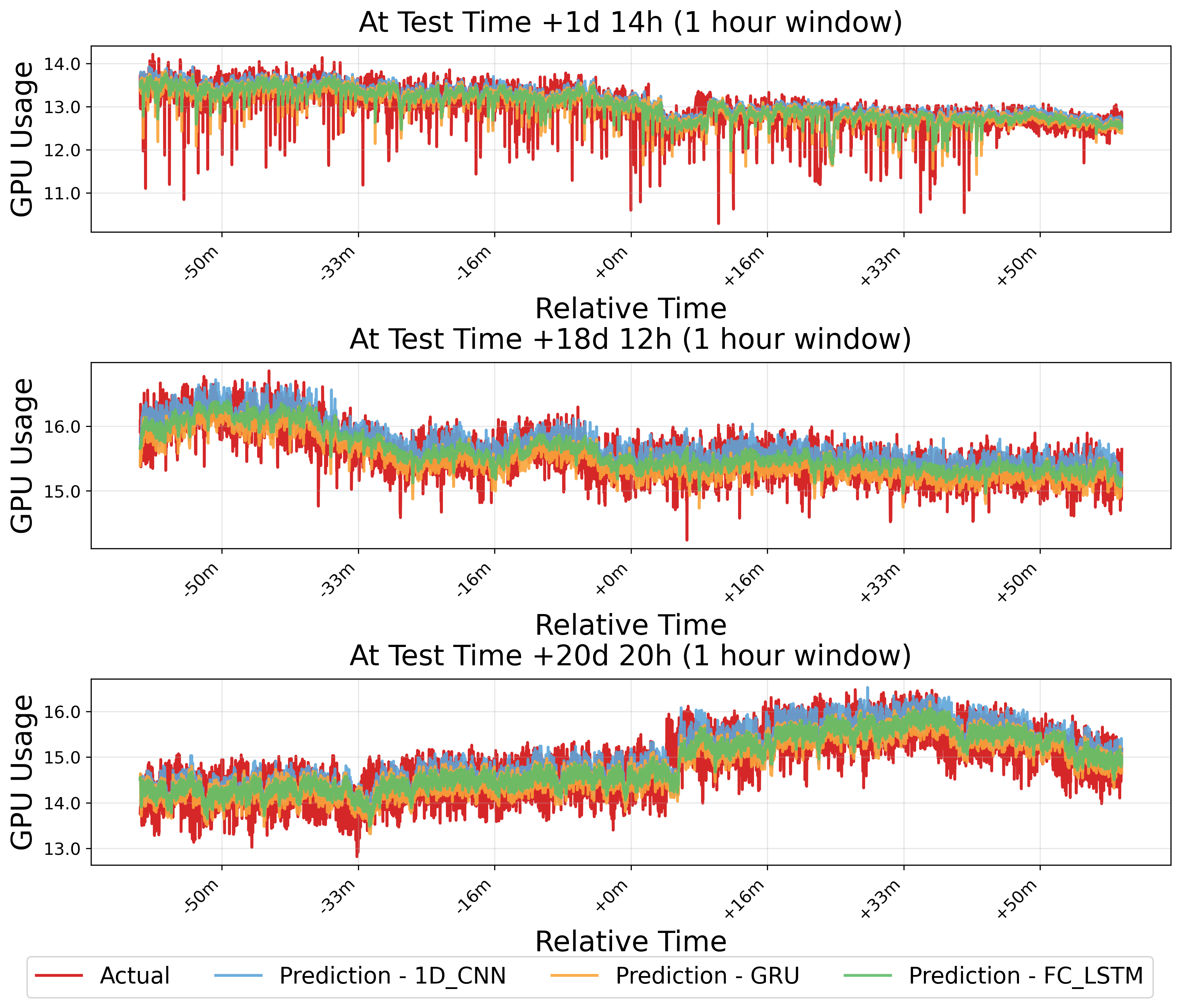}
    \caption{Prediction results (1 hour).}
    \label{fig:1_hour}    
\end{figure}

Prediction results are compared on different metrics in Table I such as RMSE, MAE, sMAPE, and R-squared. As mentioned above, LSTM, GRU and 1D-CNN are being considered for prediction methods. Fully connected LSTM has consistently achieved the best results with the lowest RMSE, MAE, sMAPE and highest R-squared value which is an indicator of robust prediction and low error. GRUs show slightly declined performance in comparison to LSTM, in terms of RMSE and R-squared error and MDB indicates slight negative bias ness in prediction results. While 1D-CNN has a relatively lower performance and noticeably more positive bias. 1-minute zoomed graphs show predictions at a finer granularity. The model’s ability to capture short-term fluctuations and variations can be observed from these graphs. 1-minute forecasting is presented in Fig.\ref{fig:1_minute}, results closely aligning with actual values. However, for sudden dips and peaks, prediction struggles to capture pattern. 10-minute zoomed plots display predictions over a longer time frame in Fig.\ref{fig:10_min}, as these graphs present a broader trend of prediction. Forecasting shows a consistent trend with real data and noticeable deviations can be seen around spikes where prediction lags the actual value. Fig.\ref{fig:1_hour} shows zoomed-in predictions vs. actual GPU consumption over a one-hour interval, depicting close tracking but with some deviations, especially in high-variability periods.

\begin{table}[h!]
\centering
\caption{Performance metrics for different models}
\begin{threeparttable}
\begin{tabular}{|c|c|c|c|c|c|}
\hline
\textbf{Model} & \textbf{RMSE} & \textbf{MAE} & \textbf{MBD} & \textbf{sMAPE} & \textbf{R\_squared} \\
\hline
FC\_LSTM & 0.5624 & 0.2916 & 0.0319 & 2.2557 & 0.9639 \\
\hline
GRU & 0.5668 & 0.3163 & -0.0433 & 2.6288 & 0.9634 \\
\hline
1D\_CNN & 0.5789 & 0.3172 & 0.1216 & 2.5014 & 0.9618 \\
\hline
\end{tabular}
\begin{tablenotes}
\scriptsize
\item \textbf{RMSE} (Root Mean Square Error): Measures the average magnitude of the errors, indicating overall accuracy.
\item \textbf{MAE} (Mean Absolute Error): Represents the average absolute difference between predicted and actual values.
\item \textbf{MBD} (Mean Bias Deviation): Measures the average bias or tendency of predictions, with positive/negative values indicating over/underestimation.
\item \textbf{sMAPE} (Symmetric Mean Absolute Percentage Error): A normalized measure of accuracy expressed as a percentage.
\item \textbf{R\_squared} (Coefficient of Determination): Indicates the proportion of the variance in the dependent variable explained by the model.
\end{tablenotes}
\end{threeparttable}
\end{table}

\subsection{Discussions and Recommendations}
This data center short-term forecasting can be considered sufficient for grid response as power generation has two main layers of control: primary control and secondary control. The primary control is the load frequency control which is dependent on inertia and governor operation and the time to response is 2-10 seconds. Secondary control is AGC which also maintains frequency and power balance in the longer term with a response time of 10-30 seconds.
While considering load-side management, this prediction analysis can provide an added advantage in the event of peak shaving with the support of battery energy storage systems (BESSs). Batteries with a capacity of several MW can be deployed in an AI-extensive system and can help in load shifting, energy arbitrage and grid resiliency. For instance, batteries can be charged during times of low electricity costs and discharged when costs are high, reducing operational expenses and huge demand spikes for data centers. This strategy can be beneficial in markets with variable electricity prices, such as those driven by real-time pricing or time-of-use rates.

\section{Conclusions}

In this paper, a short term forecasting technique is implemented on fine-grained GPU power dataset. Workload in this data-driven approach contains training of LLMs and various AI algorithms, which highlights very dynamic power consumption nature of data centers in the wild. As data centers increasingly adopt AI-intensive jobs, accurate power prediction becomes more critical. 
Future research in this direction will consider more impact factors of power consumption and include more robust algorithms like liquid neural networks for resilient and energy-efficient data center operations.

\bibliographystyle{ieeetr}
\bibliography{sample}
\end{document}